\documentclass[12pt]{article}

\usepackage{graphicx%
           ,radcor98hep%
}

\begin{document}
\def\pubnum{452}
\def\data{February, 1999}

\def\UUAABB{\hbox{%\UAB
    \vrule height0pt width2.5in
    \vbox{\hbox{\rm 
%     UAB-FT-\pubnum
     UAB-FT-\pubnum
    }\break\hbox{\data\hfill}
    \break\hbox{hep-ph/9903213\hfill} 
    \hrule height2.7cm width0pt}
   }}   
\hfill\UUAABB
\vspace{3cm}
\begin{center}
\begin{large}
\begin{bf}
$\Gamma(H^{+}\to t \bar{b})$ in the MSSM: a handle for SUSY charged 
Higgs at the Tevatron and the LHC\footnote{Talk presented at the IVth 
International Symposium on Radiative Corrections (RADCOR 98), 
Barcelona, September 8-12, 1998.  To appear in the proceedings, World 
Scientific, ed.  J.  Sol{\`a}.}\\
\end{bf}
\end{large}
\vspace{1cm}
\vspace{1cm}

J.A. COARASA, Jaume GUASCH, Joan SOL{\`A}\\

\vspace{0.25cm} 
Grup de F{\'\i}sica Te{\`o}rica\\ 
and\\ 
Institut de F{\'\i}sica d'Altes Energies\\ 
\vspace{0.25cm} 
Universitat Aut{\`o}noma de Barcelona\\
08193 Bellaterra (Barcelona), Catalonia, Spain\\
\end{center}
\vspace{0.3cm}
\begin{center}
{\bf ABSTRACT}
\end{center}
\begin{quotation}
\noindent
\hyphenation{ob-ser-va-bles com-pa-ti-bi-li-ty}
\noindent
We compute $\Gamma({\htb})$ at one-loop in the MSSM and 
show how future data at the Tevatron and/or at the LHC could be used 
to unravel the potential SUSY nature of the charged Higgs.
\end{quotation}
 
\baselineskip=6.5mm  %(FOR PREPRINT)
 
\newpage

\section{Introduction}
% The SUSY (QCD and leading electroweak) one-loop corrections 
% to the charged Higgs decay into top quark are discussed in a 
% framework in which {\tb} 
% is large and defined through $\Gamma({\Htaun})$. We show that a 
% measurement of the BR({\Htaun}), either at the TEVATRON or at 
% the LHC, 
% with a modest precision of $\sim 
% 20\%$ could be sufficient to unravel the SUSY quantum effects 
% on $\Gamma({\htb})$.
%  
%  
The charged Higgs boson can decay hadronically into several quark 
final states, and if it is sufficiently heavy it would dominantly 
decay into top and bottom quarks.  We will compute the effects of the 
leading electroweak corrections ({\SEW}) originating from large Yukawa 
couplings within the MSSM~\cite{Nilles:1984ex} as well as the SUSY-QCD 
({\SQCD}) quantum effects mediated by squarks and gluinos and shall 
compare them with the standard QCD corrections.

The vertex $H^{+}t\bar{b}$  is important not only 
for the decay under consideration but also in the production 
mechanism of the charged Higgs. Its associated production with a top 
quark would contribute to the 
cross-section for single top-quark production, whose measurement is 
one of the main goals at the next Tevatron run (Run II).  

The relevant Yukawa couplings of the charged Higgs boson with top and 
bottom quarks:
\begin{equation}
\lambda_t\equiv {h_t\over g}={m_t\over \sqrt{2}\,
M_W\,\sin{\beta}}\;\;\;\;\;,
\;\;\;\;\; \lambda_b\equiv {h_b\over g}={m_b\over \sqrt{2}
\,M_W\,\cos{\beta}}\,,
\label{eq:Yukawas}
\end{equation}
are of comparable size within the interval relevant for the Higgs 
decay
\begin{equation}
20\stackrel{\scriptstyle <}{{ }_{\sim}}\tb\stackrel{\scriptstyle <}%
{{ }_{\sim}} 50\,.
\label{eq:tansegment}
\end{equation}

\section{One-loop Corrected $\Gamma({\htb})$ in the MSSM}
\label{sec:htboneloop}

The interaction lagrangian describing the $H^{+}\,t\,\bar{b}$-vertex 
in the MSSM is:
\begin{equation}
{\cal L}_{Htb}={g\over\sqrt{2}M_W}\,H^+\,\bar{t}\, [m_t\cot\beta\,P_L
+ m_b\tan\beta\,P_R]\,b+{\rm h.c.}\,,
\label{eq:LtbH}
\end{equation}
where $P_{L,R}=1/2(1\mp\gamma_5)$ are the chiral projector operators.
From this the counterterm Lagrangian can be obtained and reads:
\begin{equation}
\delta{\cal L}_{Hbt}={g\over\sqrt{2}\,M_W}\,H^+\,\bar{t}\left[
\delta G_{L}\ m_t\,\cot\beta\,\,P_L+
\delta G_R\ m_b\,\tb\,P_R\right]\,b
+{\rm h.c.}\,,
\label{eq:LtbH2}
\end{equation}
with
\begin{eqnarray}
\delta G_{L} &=& {\delta m_t\over m_t}-{\delta v\over v}
+\frac{1}{2}\,\delta Z_{H^+}+\frac{1}{2}\,\delta Z_L^b+\frac{1}{2}
\,\delta Z_R^t
-{\delta\tb\over\tb}+\delta Z_{HW}\,\tb\,,\nonumber\\
\delta G_{R} &=& {\delta m_b\over m_b}-{\delta v\over v}
+\frac{1}{2}\,\delta Z_{H^+}+\frac{1}{2}\,\delta Z_L^t+\frac{1}{2}
\,\delta Z_R^b
+{\delta\tb\over\tb}-\delta Z_{HW}\,\cot\beta\,,\nonumber
\end{eqnarray}
and where $\delta v$ is the counterterm for 
$v=\sqrt{v_{1}^{2}+v_{2}^{2}}=\sqrt{2} M_{w}/g$, $\delta Z_{H}$ and 
$\delta Z_{HW}$ stand respectively for the charged Higgs and mixed 
$H-W$ wave-function.  The remaining are the standard wave-function and 
mass renormalization counterterms for the fermion external lines.

To fix the {\tb} counterterm we define {\tb} through 
$\Gamma({\Htaun})$.  Therefore the corrections to this decay are 
cancelled by suitable counterterms.  From this condition and 
parametrizing the different contributions at one loop to {\htb} in 
terms of two form factors $H_{L}$, $H_{R}$ the one-loop corrected 
{\htb} vertex in the MSSM is:
\begin{equation}
\Lambda = {i\,g\over\sqrt{2}\,M_W}
\,\left[m_t\,\cot\beta\,(1+G_L)\,P_L
 + m_b\,\tb\,(1+\Lambda_R)\,P_R\right]\,,
\label{eq:AtbH}
\end{equation}
where
\begin{eqnarray}
G_L & = & H_{R}+{\delta m_t\over m_t}
+\frac{1}{2}\,\delta Z_L^b+\frac{1}{2}\,\delta Z_R^t-
\Delta_{\tau}\nonumber\\
& - & {\delta v^2\over v^2}+\delta Z_{H^+}+(\tb-\cot\beta)\,
\delta Z_{HW}
 \,,\nonumber\\
G_{R} &=& H_{L}+{\delta m_b\over m_b}
+\frac{1}{2}\,\delta Z_L^t+\frac{1}{2}\,\delta Z_R^b
+\Delta_{\tau}\,,
\label{eq:lambdaLR}
\end{eqnarray}
and $\Delta_{\tau}$ is a {\Htaun} process dependent  
contribution coming from our {\tb} definition condition.

In the following we will describe the relevant electroweak one-loop 
supersymmetric diagrams entering the amplitude of {\htb} in the MSSM.  
At one-loop, we have the diagrams exhibited in 
Figs.~\ref{fig:htbsusy}-\ref{fig:htbhiggs} apart from the ones 
entering the calculation of the different 
counterterms~\cite{Coarasa:1996qa}.  The computation of the one-loop 
diagrams requires to use the full structure of the 
MSSM~\cite{Nilles:1984ex} Lagrangian.  The explicit form of the most 
relevant pieces of this Lagrangian, together with the necessary SUSY 
notation, is provided in the Appendix.

\fightbsusy

\subsection{SUSY vertex diagrams}

Following the labelling in Fig.~\ref{fig:htbsusy} we find the form 
factors $H_{L}$, $H_{R}$.
\begin{itemize}
        \item {Diagram $(V_{S0})$:}
        Using the convention that lower indices are 
        summed over, whereas upper indices 
        are just for notational convenience one finds:
        \begin{eqnarray} 
                H_L&=8\pi\alpha _s\,i C_F
                \displaystyle{\frac{G_{ab}^*}{m_t\cot\beta}} &
                [R^{(t)}_{1 b}R^{(b)*}_{1 a}(\Coo-\Cot)m_t+ 
                R^{(t)}_{2 b}R^{(b)*}_{2 a}\Cot m_b\nonumber\\
                &&+ R^{(t)}_{2 b}R^{(b*)}_{1 a}\Cz \mg]\,,\nonumber\\
                H_R&=8\pi\alpha _s\,i C_F
                \displaystyle{\frac{G_{ab}^*}{m_b \tan\beta}} &
                [R^{(t)}_{2 b}R^{(b)*}_{2 a}(\Coo-\Cot)m_t+ 
                R^{(t)}_{1 b}R^{(b)*}_{1 a}\Cot m_b\nonumber\\
                &&+R^{(t)}_{1 b}R^{(b)*}_{2 a}\Cz \mg] \,,
                \label{V1sqcd}
        \end{eqnarray} 
        where the various $3$-point functions are as in 
        ref.~\cite{Coarasa:1996qa}, so that, in eq.~(\ref{V1sqcd}) the 
        C-functions must be evaluated with arguments:
        \[
                %\label{V0Cs}
                C_{*}=C_{*}\left(p,p',m_{\tilde{g}},
		m_{\tilde{t}_b},m_{\tilde{b}_a}\right)\,,
        \]
        and $C_F=(N_C^2-1)/2N_C=4/3$ is the colour factor.

        \item {Diagram $(V_{S1})$:}
        Making use of the coupling matrices of eqs.~(\ref{V1Apm}) and 
        (\ref{eq:QLQR}) we introduce the shorthands
        \[
                \label{V1Apmdef}
                \Apm\equiv\Apmit\ \ \mbox{and}\ \ \Apmz\equiv\Apmat\,,
        \]
        and define the combinations (omitting indices also 
	for $\QaiL,\QaiR$)
\begin{eqnarray}
  \label{V1matrices}
    \RRLo =\cbt\Apc\HRc\Amz\,,&\ \ \ 
   &\RRRo =\cbt\Amc\HRc\Amz\,,\nonumber\\ %A(1),E(1)
    \LRLo =\cbt\Apc\HRc\Apz\,,&\ \ \ 
   &\LRRo =\cbt\Amc\HRc\Apz\,,\nonumber\\ %B(1),F(1)
    \RLLo =\sbt\Apc\HLc\Amz\,,&\ \ \ 
   &\RLRo =\sbt\Amc\HLc\Amz\,,\nonumber\\ %C(1),G(1)
    \LLLo =\sbt\Apc\HLc\Apz\,,&\ \ \ 
   &\LLRo =\sbt\Amc\HLc\Apz\,.            %D(1),H(1)
\end{eqnarray}

The contribution from diagram $(V_{S1})$ to the form factors $H_{R}$ 
and $H_{L}$ is:
\begin{eqnarray}
  \label{V1FF}
  H_{R}&=&M_L\left[\LLRo\Czt+\right.\nonumber\\
  +&&\mb\,\left(\mt\RRLo+\ma\LRLo+\mb\LLRo+\mi\LLLo\right)\Cot
  \nonumber\\
  +&&\mt\,\left(\mt\LLRo+\ma\RLRo+\mb\RRLo+\mi\RRRo\right)
     \left(\Coo-\Cot\right)\nonumber\\
  +&&\left.\left(\mt\mb\RRLo+\mt\mi\RRRo+\ma\mb\LRLo+\mi\ma\LRRo
  \right)
     \Cz\right]\,,\nonumber\\
  H_{L}&=&M_R\left[\RRLo\Czt+\right.\nonumber\\
  +&&\mb\,\left(\mt\LLRo+\ma\RLRo+\mb\RRLo+\mi\RRRo\right)\Cot
  \nonumber\\
  +&&\mt\,\left(\mt\RRLo+\ma\LRLo+\mb\LLRo+\mi\LLLo\right)
     \left(\Coo-\Cot\right)\nonumber\\
  +&&\!\!\!\!\left.\left(\mt\mb\LLRo+\mt\mi\LLLo+
  \ma\mb\RLRo+\mi\ma\RLLo\right)
     \Cz\right] ,
\end{eqnarray}
where the overall coefficients $M_L$ and $M_R$ are:
\begin{equation}
  \label{MLMR}
  M_L=-\frac{ig^2\mw}{\mb\tb}\ \ \ \ M_R=-\frac{ig^2\mw}{\mt\ctb}\,.
\end{equation}
%The notation for the various $3$-point functions is summarized in 
%Appendix~\ref{sec:pointfunct}. 
In eq.~(\ref{V1FF}) the C-functions must be evaluated with arguments:
\begin{equation}
  \label{V1Cs}
  C_{*}=C_{*}\left(p,p',\msta,\ma,\mi\right)\,.
\end{equation}

\item {Diagram $(V_{S2})$:}
For this finite diagram  we use the matrices on
eqs.~(\ref{V1Apm}) and (\ref{gdef}), and introduce the shorthands
\[
%  \label{V2Apmdef}
   \Apmb\equiv A_{\pm b\alpha}^{(b)}\ \ \mbox{and}\ \ \Apmt\equiv
   \Apmat\,,
\]
to define the products of coupling matrices
\begin{eqnarray*}
  \label{V2matrices}
  \RLt =\Gba\Apbc\Amt\,,&\ \ \ &\RRt =\Gba\Ambc\Amt\,,\nonumber\\ 
  %A(2),C(2)
  \LLt =\Gba\Apbc\Apt\,,&\ \ \ &\LRt =\Gba\Ambc\Apt\,.            
  %B(2),D(2)
\end{eqnarray*}
The contribution to the form factors $H_{R}$ and $H_{L}$ from this 
diagram is
\begin{eqnarray*}
  \label{V2FF}
  H_{R}&=&\frac{M_L}{2\mw}\left[
  \mb\LLt\Cot+\mt\RRt\left(\Coo-\Cot\right)-\ma\LRt\Cz\right]\,,
  \nonumber\\
  H_{L}&=&\frac{M_R}{2\mw}\left[
  \mb\RRt\Cot+\mt\LLt\left(\Coo-\Cot\right)-\ma\RLt\Cz\right]\,,
\end{eqnarray*}
the coefficients $M_L$, $M_R$ being those of eq.~(\ref{MLMR}) and
\[
%  \label{V2Cs}
  C_{*}=C_{*}\left(p,p',\ma,\msta,\msbb\right)\,.
\]

\item {Diagram $(V_{S3})$:}
For this diagram  we will need
\[
%  \label{V3Apmdef}
  \Apm\equiv\Apmib\ \ \mbox{and}\ \ \Apmz\equiv\Apmab\,,
\]
and again omitting indices we shall use
\begin{eqnarray}
  \label{V3matrices}
    \RRLth =\cbt\Apzc\HRc\Am\,,&\ \ \ 
   &\RRRth =\cbt\Amzc\HRc\Am\,,\nonumber\\%A(3),E(3)
    \LRLth =\cbt\Apzc\HRc\Ap\,,&\ \ \ 
   &\LRRth =\cbt\Amzc\HRc\Ap\,,\nonumber\\%B(3),F(3)
    \RLLth =\sbt\Apzc\HLc\Am\,,&\ \ \ 
   &\RLRth =\sbt\Amzc\HLc\Am\,,\nonumber\\%C(3),G(3)
    \LLLth =\sbt\Apzc\HLc\Ap\,,&\ \ \ 
   &\LLRth =\sbt\Amzc\HLc\Ap\,.           %D(3),H(3)
\end{eqnarray}
{}From these definitions the contribution of diagram $(V_{S3})$ to the 
form factors can be obtained by performing the following changes in 
that of diagram $(V_{S1})$, eq.~(\ref{V1FF}):
\begin{itemize}
\item Everywhere in eqs.~(\ref{V1FF}) and (\ref{V1Cs}) replace 
$\mi\leftrightarrow\ma$ and $\msta\leftrightarrow\msba$.
\item Replace in eq.~(\ref{V1FF}) couplings from (\ref{V1matrices}) 
with those of (\ref{V3matrices}).
\item Include a global minus sign.
\end{itemize}
\end{itemize}

\subsection{Higgs vertex diagrams}

For the contributions arising from the exchange of virtual
Higgs particles
and Goldstone bosons in the Feynman gauge,
Fig.~\ref{fig:htbhiggs}, we write the formula for the form factors
by giving the value of the
overall coefficient $N$ 
and the arguments of the corresponding $3$-point functions.
\begin{itemize}
\item {Diagram $(V_{H1})$:}
\begin{eqnarray*}
  H_{R}&=&N\,[\mbs (\Cot-\Cz)+\mts \ctbs (\Coo-\Cot) ]\,,\\
  H_{L}&=&N\mbs [\Cot-\Cz +\tbs (\Coo-\Cot) ] \,,\\
  N&=&\mp\frac{ig^2}{2}\left(1-\frac{\{\mHs,\mhs\}}{2\mws}\right)
        \frac{\{\ca , \sa \}}{\cbt}
	\{{\scriptstyle{\cbma , \sbma}}\}\,, \\
  C_{*}&=&C_{*}\left(p,p',\mb,\mHp,\{\mH,\mh\}\right)\,.
\end{eqnarray*}

\fightbhiggs
\item {Diagram $(V_{H2})$:}
\begin{eqnarray*}
  H_{R}&=&N\ctb[\mts (\Coo -\Cot )+\mbs(\Cz -\Cot )]\,,\\
  H_{L}&=&N\mbs\tb (2\Cot -\Coo -\Cz ) \,,\\
  N&=&\frac{ig^2}{4} \frac{\{\ca , \sa \}}{\cbt}
  \{{\scriptstyle{\sbma  , \cbma}}\}
        \left(\frac{\mHps}{\mws}-\frac{\{\mHs  , \mhs \}}
{\mws}\right)\,, \\
  C_{*}&=&C_{*}\left(p,p',\mb,\mw,\{\mH,\mh\}\right)\,.
\end{eqnarray*}

\item {Diagram $(V_{H3})$:}
\begin{eqnarray*}
  H_{R}&=&N\mts [\ctbs\Cot +\Coo -\Cot -\Cz ] \,,\\
  H_{L}&=&N\,[\mbs\tbs\Cot  +\mts (\Coo -\Cot-\Cz )]\,, \\
  N&=&-\frac{ig^2}{2}\frac{\{\sa , \ca \}}{\sbt}
  \{{\scriptstyle{\cbma  , \sbma}}\}
        \left(1-\frac{\{\mHs  ,\mhs \}}{2\mws} \right) \,,\\
  C_{*}&=&C_{*}\left(p,p',\mt,\{\mH,\mh\},\mHp\right)\,.
\end{eqnarray*}

\item {Diagram $(V_{H4})$:}
\begin{eqnarray*}
  H_{R}&=&N\mts (2\Cot-\Coo+\Cz)\ctb \,,\\
  H_{L}&=&N\,[-\mbs\Cot +\mts (\Coo-\Cot-\Cz)]\tb\,, \\
  N&=&\mp\frac{ig^2}{4}\frac{\{\sa , \ca \}}{\sbt}
  \{{\scriptstyle{\sbma  , \cbma}}\}
        \left(\frac{\mHps}{\mws}-\frac{\{\mHs  ,\mhs \}}
{\mws}\right) \,,\\
  C_{*}&=&C_{*}\left(p,p',\mt,\{\mH,\mh\},\mw\right)\,.
\end{eqnarray*}

\item {Diagram $(V_{H5})$:}
\begin{eqnarray*}
  H_{R}&=&N\,[\mbs(\Cot+\Cz)+\mts(\Coo-\Cot)]\,, \\
  H_{L}&=&N \mbs \tbs (\Coo+\Cz)\,, \\
  N&=&-\frac{ig^2}{4}\left(\frac{\mHps}{\mws}-\frac{\mAs}
{\mws}\right)\,, \,\,\,
  C_{*}=C_{*}\left(p,p',\mb,\mw,\mA\right)\,.
\end{eqnarray*}

\item {Diagram $(V_{H6})$:}
\begin{eqnarray*}
  H_{R}&=& N\mts\ctbs (\Coo+\Cz)\,, \\
  H_{L}&=& N\,[\mbs\Cot +\mts (\Coo-\Cot+\Cz)]\,, \\
  N&=&-\frac{ig^2}{4}\left(\frac{\mHps}{\mws}-\frac{\mAs}
{\mws}\right) \,,\,\,\,
  C_{*}=C_{*}\left(p,p',\mt,\mA,\mw\right)\,.
\end{eqnarray*}

\item {Diagram $(V_{H7})$:}
\begin{eqnarray*}
  H_{R}&=N&\,[(2\mbs\Coo+\Czt+2(\mts-\mbs)(\Coo-\Cot ))\ctbs \\
        &&+2\mbs(\Coo+2\Cz)]\mts\,, \\
  H_{L}&=N&\,[(2\mbs\Coo+\Czt+2(\mts-\mbs)(\Coo-\Cot ))\tbs \\
        &&+2\mts(\Coo+2\Cz)]\mbs\,, \\
  N&=&\pm\frac{ig^2}{4\mws}\frac{\sa \ca}{\sbt \cbt}\,, \\
  C_{*}&=&C_{*}\left(p,p',\{\mH,\mh\},\mt,\mb\right)\,.
\end{eqnarray*}

\item {Diagram $(V_{H8})$:}
\begin{eqnarray*}
  H_{R}&=& N\mts\ctbs\,\Czt\,,  \\
  H_{L}&=& N\mbs\tbs\,\Czt\,,   \\
  N&=&\mp \frac{ig^2}{4\mws}\,, \,\,\,
  C_{*}=C_{*}\left(p,p',\{\mA,\mz\},\mt,\mb\right)\,.
\end{eqnarray*}

\end{itemize}

\section{Numerical Analysis}

The relevant MSSM parameter region where we have carried out the 
numerical analysis has been obtained in 
accordance~\cite{Coarasa:1997ky} with the CLEO data~\cite{Alam:1995aw} 
on radiative $\bar{B}^0$ decays at $2\,\sigma$, imposing also that 
non-SM contributions to the $\rho$-parameter be tempered by the 
relation
\begin{equation}
\delta\rho_{\rm new}\leq 0.003\,.
\label{eq:deltarho}
\end{equation}
and having checked that the known necessary conditions for the 
non-existence of colour-breaking minima are fulfilled.  Where the 
charged Higgs
boson mass %$M_H$ 
has to be fixed, we have chosen the value $M_H=250\,GeV$ within the 
range:
\begin{equation}
m_t\stackrel{\scriptstyle <}{{ }_{\sim}} M_H\stackrel{\scriptstyle %
<}{{ }_{\sim}} 300\,GeV\,.
\label{eq:interval}
\end{equation}
This window is especially significant in that the CLEO 
measurements~\cite{Alam:1995aw} of $BR(b\rightarrow s\,\gamma)$ 
forbid most of this domain within the context of a generic $2$HDM.  
However, within the MSSM the mass interval (\ref{eq:interval}) is 
perfectly consistent provided that relatively light stop and charginos 
($\stackrel{\scriptstyle <}{{ }_{\sim}} 200\,GeV$) occur.  
Nevertheless, we shall also explicitly show the evolution of our 
results with $M_H$.

\fightbbrtau
We set out by looking at the branching ratio of 
$H^+\rightarrow\tau^+\,\nu_{\tau}$ (Cf.  Fig.\ref{fig:htbhbrtau}).  
Even though the partial width of this process does not get 
renormalized, its branching ratio is seen to be very much sensitive to 
the MSSM corrections to $\Gamma(H^+\rightarrow t\,\bar{b})$.  Taking 
the standard QCD-corrected branching ratio (central curve in that 
figure) as a fiducial quantity then 
$BR(H^+\rightarrow\tau^+\,\nu_{\tau})$ undergoes an effective MSSM 
correction of order $\pm (40-50)\%$.  The sign of this effect is given 
by the sign of $\mu$.

Moreover, for large $\tb$ as in eq.(\ref{eq:tansegment}), 
$BR(H^{+}\rightarrow\tau^{+}\nu_{\tau})$ may achieve rather high 
values ($10-50\%$) for Higgs masses in the interval 
(\ref{eq:interval}), and it never decreases below the $5-10\%$ level 
in the whole range.  Therefore, a handle for $\tb$ measurement is 
always available from the Higgs $\tau$-channel and so also an 
opportunity for discovering quantum SUSY signatures on $\Gamma 
(H^+\rightarrow t\,\bar{b})$.  As for the other $H^\pm$-decays, we 
note that the potentially important mode 
$H^{+}\rightarrow\tilde{t}_i\,\bar{\tilde{b}}_j$ does not play any 
role in our case since (for reasons to be clear below) we are mainly 
led to consider bottom-squarks heavier than the charged Higgs.  
Moreover, the $H^+\rightarrow W^+\,h^0$ decay which is sizable enough 
at low $\tb$ becomes extremely depleted at high $\tb$ 
~\cite{Jimenez:1996wf}.  Finally, the decays into charginos and 
neutralinos, $H^+\rightarrow \chi^+_i\,\chi_{\alpha}^0$, are not 
$\tb$-enhanced and remain negligible.  Thus we find an scenario where 
$H^+\rightarrow t\,\bar{b}$ and $H^{+}\rightarrow\tau^{+}\,\nu_{\tau}$ 
are the only relevant decay modes.

In order to assess the impact of the electroweak effects, we find a 
typical set of inputs such that the {\SQCD} and {\SEW} outputs are of 
comparable size.

In Figs.\ref{fig:htbdpos}a-\ref{fig:htbdpos}b we display the 
correction $\delta$ defined with respect to the tree level width 
$\Gamma =\Gamma({\htb})$:
\begin{equation}
    \delta = {\Gamma -\Gamma^{(0)}\over \Gamma^{(0)}}
    \label{eq:eq:deltaalpha}
\end{equation}%
as a function respectively of $\mu<0$ and $\tb$ for fixed values of 
the other parameters (within the $b\rightarrow s\,\gamma$ allowed 
region).  Remarkably, in spite of the fact that all sparticle masses 
are beyond the scope of LEP$\,200$ the corrections are fairly large.  
We have individually plot the {\SEW}, {\SQCD}, standard QCD and total 
MSSM effects.  The Higgs-Goldstone boson corrections are isolated only 
in Fig.\ref{fig:htbdpos}b just to make clear that they add up 
non-trivially to a very tiny value in the whole range 
(\ref{eq:tansegment}), and only in the small corner $\tb<1$ they can 
be of some significance.  In{\fightbdpos} 
Figs.\ref{fig:htbdpos}c-\ref{fig:htbdpos}d we render the various 
corrections as a function of the relevant squark masses.  For 
$m_{\tilde{b}_1}\stackrel{\scriptstyle <}{{ }_{\sim}} 200\,GeV$ we 
observe (Cf.  Fig.\ref{fig:htbdpos}c) that the {\SEW} contribution is 
non-negligible ($\delta_{SUSY-EW}\simeq +20\%$) but the {\SQCD} loops 
induced by squarks and gluinos are by far the leading SUSY effects 
($\delta_{SUSY-QCD}> 50\%$) -- the standard QCD correction staying 
invariable over $-20\%$ and the standard EW correction (not shown) 
being negligible.  In contrast, for larger and larger 
$m_{\tilde{b}_1}>300\,GeV$, say $m_{\tilde{b}_1}=400$ or $500\,GeV$, 
and fixed stop mass at a moderate value $m_{\tilde{t}_1}=150\,GeV$, 
the {\SEW} output is longly sustained whereas the {\SQCD} one steadily 
goes down.  However, the total SUSY pay-off adds up to about $+40\%$ 
and the net MSSM yield still reaches a level around $+20\%$, i.e.  of 
equal value but opposite in sign to the conventional QCD result.  This 
would certainly entail a qualitatively distinct quantum signature.

We stress that the main parameter to decouple the {\SQCD} correction 
is the lightest sbottom mass, rather than the the gluino 
mass~\cite{Jimenez:1996wf}, with which the decoupling is very slow.  
For this reason, since we wished to probe the regions of parameter 
space where these electroweak effects are important, the direct SUSY 
decay $H^{+}\rightarrow\tilde{t}_i\,\bar{\tilde{b}}_j$ mentioned above 
is blocked up kinematically and plays no role in our analysis.  On the 
other hand, the {\SEW} output is basically controlled by the lightest 
stop mass, as it is plain in Fig.\ref{fig:htbdpos}d, where we vary it 
in a range past the LEP$\,200$ threshold.

We have also checked that in the alternative $\mu>0$, $A_t<0$ 
scenario, the {\SQCD} correction is negative but it is largely 
cancelled by the {\SEW} part, which stays positive, so that the total 
$\delta_{MSSM}$ is negative and larger (in absolute value) than the 
standard QCD correction.

% \fightbleading
% 
% \fightbleadingtau

\section{Conclusions}

We have presented a fairly complete treatment of the supersymmetric 
quantum effects ({\SQCD} and {\SEW}) on the decay width of 
$H^+\rightarrow t\,\bar{b}$ and have put forward evidence that they 
could be sizable enough to seriously compete with the ordinary QCD 
corrections.  Consequently, they can either reinforce the conventional 
QCD corrections or counterbalance them, and even reverse their sign.  
This should be helpful to differentiate $H^+$ from alternative charged 
pseudoscalar decays leading to the same final states.

Our computation shows that
these effects are compatible with CLEO data from
low-energy $B$-meson phenomenology.
We confirm that also in the constrained case
the {\SQCD} effects are generally very important
(typically between $10\%-50\%$), slowly decoupling and of both 
signs~\cite{Jimenez:1996wf}.
However, we have exemplified an scenario with sparticle masses above
the LEP$\,200$ discovery range where the SUSY electroweak corrections
triggered by large Yukawa couplings can be comparable to the
{\SQCD} effects. In this context the total SUSY correction
remains fairly large --around $+(30-50)\%$-- with a $\sim 50\%$ 
component from electroweak supersymmetric origin.
This situation occurs for 
\begin{itemize}
    \item large $\tb$ ($>20$),

    \item huge sbottom masses ($> 300\, GeV$) and 

    \item relatively light stop and charginos ($100-200\,GeV$).
\end{itemize}
If the charged Higgs mass lies in the 
intermediate window (\ref{eq:interval}),
a chance is still left for Tevatron to produce a charged Higgs
heavier than the top quark by means of
``charged Higgsstrahlung'' off top and bottom quarks.
Should, however, a heavier $H^\pm$ exist outside the window 
(\ref{eq:interval}), the LHC could continue the searching task mainly 
from gluon-gluon fusion where again $H^\pm$ is produced in association 
with the top quark.

The upshot is that the whole range of charged Higgs masses up to about 
$1\,TeV$ could be probed and, within the present renormalization 
framework, its potential supersymmetric nature be unravelled through a 
measurement of $\Gamma (H^+\rightarrow t\,\bar{b})$ with a modest 
precision of $\sim 20\%$.  Alternatively, one could look for indirect 
SUSY quantum effects on the branching ratio of $H^+\rightarrow 
\tau^+\,\nu_{\tau}$ by measuring this observable to within a similar 
degree of precision.

\section*{Acknowledgments}

The work of J.G.  has been financed by a grant of the Comissionat per 
a Universitats i Recerca, Generalitat de Catalunya.  This work has 
also been partially supported by CICYT under project No.  AEN95-0882.

\section*{Appendix}
\label{sec:susyapp}

Within the context of the MSSM~\cite{Nilles:1984ex} we need two Higgs 
superfield doublets
\begin{equation}
 \hat{H}_1= \left(\begin{array}{c}
\hat{H}_1^0 \\ \hat{H}_1^{-}
\end{array} \right)
 \;\; \;\;\;,\;\;\;\;\;
 \hat{H}_2= \left(\begin{array}{c}
\hat{H}_2^{+} \\ \hat{H}_2^0 
\end{array} \right)\,,  
\end{equation}
with weak hypercharges $Y_{1,2}=\mp 1$.  The (neutral components of 
the) corresponding scalar Higgs doublets give mass to the down (up) 
-like quarks through the VEV $<H^0_1>=v_1$ ($<H^0_2>=v_2$).
This is seen from the structure of the MSSM 
superpotential
\begin{equation}
\hat{W}=\epsilon_{ij}[h_b\,\hat{H}_1^i\hat{Q}^j\hat{D}+
h_t\hat{H}_2^j\hat{Q}^i\hat{U}-\mu\hat{H}_1^i\hat{H}_2^j]\,,
\label{eq:W}
\end{equation}
where we have singled out only the Yukawa couplings of the third 
quark-squark generation, $(t,b)-(\tilde{t},\tilde{b})$, as a generic 
generation of chiral matter superfields $\hat{Q}$, $\hat{U}$ and 
$\hat{D}$.
Their respective scalar (squark) components are: 
\begin{equation}
\tilde{Q}=\left(\begin{array}{c}
\tilde{t'}_L \\ \tilde{b'}_L
\end{array} \right) \; \;\; ,\;\;\;\tilde{U}=\tilde{t'}_R^{*}
\; \;\; ,\;\;\;\tilde{D}=\tilde{b'}_R^{*}\,,
\label{eq:QUD} 
\end{equation}
with weak hypercharges $Y_Q=+1/3$, $Y_U=-4/3$ and $Y_D=+2/3$. 
The primes in (\ref{eq:QUD}) denote the fact that 
$\tilde{q'}_a=\{\tilde{q'}_L, \tilde{q'}_R\}$ are weak-eigenstates, 
not mass-eigenstates.
The ratio
\begin{equation}
  \tan\beta={v_2\over v_1}\,,
  \label{eq:v2v1}
\end{equation}
is a most relevant parameter throughout our analysis.

We briefly describe the necessary SUSY formalism:
\begin{itemize}
\item The fermionic partners of the weak-eigenstate gauge bosons and
  Higgs bosons are called gauginos, $\tilde{B}$, $\tilde{W}$, and
  higgsinos, $\tilde{H}$, respectively. From them we construct the
  mass-eigenstates, so-called charginos and neutralinos, by
  forming the following three sets of two-component Weyl
  spinors:
  \begin{equation}
    \Gamma_i^{+} = \{-i\tilde{W}^{+}, \tilde{H_2^{+}}\}\ , \;\;\;
    \Gamma_i^{-} = \{-i\tilde{W}^{-}, \tilde{H_1^{-}}\}\ , \nonumber
  \end{equation}
  \begin{equation}
    \Gamma_{\alpha}^{0} =
    \{-i\tilde{B^{0}}, -i\tilde{W_3^{0}}, \tilde{H_2^0}, 
    \tilde{H_1^0}\}\ , 
    \nonumber  
  \end{equation}
  which get mixed up when the neutral Higgs fields acquire 
  nonvanishing 
  VEV's, and diagonalizing the resulting ``ino'' mass Lagrangian
 \begin{eqnarray}
  {\cal L}_M&=&-<\Gamma^{+}|
              \pmatrix{
                  M                       &  \sqrt{2} M_W \cbtt \cr
                  \sqrt{2} M_W \sbtt &  \mu 
              }
%{\cal M}
              |\Gamma^{-}>\nonumber\\
              -{1\over2}&
              <\Gamma^0|&
%{\cal M}^0
          \! \! \! \! \pmatrix{
M'     \!          & 0    \! \!             
& M_Z\sbtt s_W\! & -M_Z\cbtt s_W\!\cr
0      \!          & M    \!  \!            
&-M_Z\sbtt c_W\! & M_Z\cbtt c_W \!\cr
M_Z\sbtt s_W \! & -M_Z\sbtt c_W \! \! 
&  0\!                & -\mu\!\cr
-M_Z\cbtt s_W\! & M_Z\cbtt c_W  \! \! 
&  -\mu\!             & 0\!
              } 
              |\Gamma^0>\nonumber\\
              +&h.c.&
              \label{eq:CNMM}
  \end{eqnarray}
  where we remark the presence of the parameter $\mu$ introduced above 
  and of the soft SUSY-breaking Majorana masses $M$ and $M'$, usually 
  related as $M'/ M=(5/3)\,\tan^2{\theta_{W}}$, and where $\cbtt=\cbt$ 
  and $\sbtt=\sbt$.  The corresponding
  mass-eigenstates\footnote{We use the following notation:
    first Latin indices a,b,...=1,2 are reserved for sfermions,
    middle Latin indices i,j,...=1,2 for charginos, and first Greek
    indices $\alpha, \beta,...=1,\ldots ,4$ for neutralinos. } 
  are:
  \begin{equation}
    \Psi_i^{+}= \pmatrix{
                U_{ij}\Gamma_j^{+} \cr V_{ij}^{*}\bar{\Gamma}_j^{-}
                }
              \; \;\; \;\;,\;\;\;\;\;
              \Psi_i^{-}= C\bar{\Psi_i}^{-T} =\pmatrix{
                  V_{ij}\Gamma^{-}_j \cr U_{ij}^{*}\bar{\Gamma}_j^{+} 
                }\ ,  
              \label{eq:cinos}
              \nonumber
  \end{equation}
  and
  \begin{equation}
    \Psi_{\alpha}^0= \pmatrix{
                N_{\alpha\beta}\Gamma_{\beta}^0 \cr 
                N_{\alpha\beta}^{*}\bar{\Gamma}_{\beta}^0
                } =  
              C\bar{\Psi}_{\alpha}^{0T}\ ,
              \label{eq:ninos} 
              \nonumber
  \end{equation}
  where the matrices $U,V,N$ are defined through
  \begin{equation}
    U^{*}{\cal M}V^{\dagger}=diag\{M_1,M_2\}\;,\;\;\;
    N^{*}{\cal M}^0N^{\dagger}=diag\{M_1^0,\ldots, M_4^0\}\ . 
\label{eq:UVN}
  \end{equation}    

  Among the gauginos we also have the strongly interacting gluinos, 
  $\sg^r$ $(r=1,\ldots,8)$,
  which are the fermionic partners of the gluons.
\item As for the scalar partners of quarks and leptons,
they are called squarks,
  $\tilde{q}$, and sleptons, $\tilde{l}$, respectively.
Again we will use the third quark-squark 
generation $(t,b)-(\tilde{t},\tilde{b})$ as a
   generic fermion-sfermion generation.  The squark
  mass-eigenstates,
  $\tilde{q}_a=\{\tilde{q}_1, \tilde{q}_2\}$, if we neglect 
  intergenerational mixing, are obtained from the weak-eigenstate ones 
  $\tilde{q'}_a=\{\tilde{q'}_1\equiv \tilde{q}_L,\,\, \tilde{q'}_2
  \equiv
  \tilde{q}_R\}$, through 
\begin{eqnarray}
 \tilde{q'}_a&=&\sum_{b}
     R_{ab}^{(q)}\tilde{q}_b,\nonumber\\ R^{(q)}&
     =&\pmatrix{ \cos{\theta_q} & -\sin{\theta_q} \cr 
         \sin{\theta_q} & \cos{\theta_q}}
\;\;\;\;\;\;
       (q=t, b)\,.
\label{eq:rotation}
\end{eqnarray}
 The rotation matrices in
    (\ref{eq:rotation}) diagonalize the corresponding stop and sbottom
    mass matrices:
  \begin{equation}
    {\cal M}_{\tilde{q}}^2 =\pmatrix{
      M_{\tilde{q}_L}^2+m_q^2+\cos{2\beta}(\TqL-\Qq s_W^2) 
      M_Z^2 \ \ \ \ \ \  \ \ 
      m_q M_{LR}^q\cr 
      m_q M_{LR}^q \ \  \ \ \ \  \ \ 
      M_{\tilde{q}_R}^2+m_q^2+ \cos{2\beta}\,\Qq s_W^2 M_Z^2
      }\,,
    \label{eq:stopmatrix}
    \label{eq:sbottommatrix}
    \end{equation}
\begin{equation}
R^{(q)\dagger} {\cal M}_{\tilde{q}}^2 R^{(q)}=
      diag\{m_{\tilde{q}_2}^2,m_{\tilde{q}_1}^2\}
\ \ \ \ \ (m_{\tilde{q}_2}\geq m_{\tilde{q}_1})\,, 
\end{equation}    
    with $\TqL$ the third component of weak isospin, $Q$ the electric
    charge, and
 $M_{{\tilde{q}}_{L,R}}$ the soft SUSY-breaking squark 
 masses. (By
    $SU(2)_L$-gauge invariance, we must have
    $M_{\tilde{t}_L}=M_{\tilde{b}_L}$, whereas $M_{{\tilde{t}}_R}$,
    $M_{{\tilde{b}}_R}$ are in general independent parameters.)
The mixing angle on eq.(\ref{eq:rotation}) is given by
\begin{equation}
 \tan2{\theta_q} = {2\,m_q\,M_{LR}^q\over
M_{\tilde{q}_L}^2-M_{\tilde{q}_R}^2+\cos{2\beta}(\TqL-2\Qq s_W^2) 
M_Z^2}\,,
       \label{eq:thetarot}
\end{equation}
where
\begin{equation}
 M_{LR}^t=A_t-\mu\cot\beta\,,\ \ \ \ \  
 M_{LR}^b=A_b-\mu\tan\beta\,,
\end{equation}
are, respectively, the stop and sbottom off-diagonal mixing terms
on eq.(\ref{eq:sbottommatrix}), and $A_{t,b}$
are the trilinear soft SUSY-breaking parameters. 

The charged slepton mass-eigenstates can be obtained in a similar way.
  \end{itemize}

\begin{itemize}
\item{\bf fermion--sfermion--(chargino or neutralino)}

After translating the 
allowed quark-squark-higgsino/gaugino interactions into the
mass-eigenstate basis, one finds 
\begin{eqnarray}
  \label{LcqsqLR}
  {\cal L}_{\Psi q\tilde{q}}&=&
  g\,\sum_{a=1,2}\sum_{i=1,2}\left( 
    -\sta^*\cbim\left(\Apit\PL+\Amit\PR\right)\,b\right.\nonumber\\
    &&\phantom{g\,\sum_{a=1,2}\sum_{i=1,2}}
    -\left.\sba^*\cbip\left(\Apib\PL+\Amib\PR\right)\,t 
  \right) \nonumber\\
   +&\displaystyle{\frac{g}{\sqrt{2}}}&
    \sum_{a=1,2}\sum_{\alpha=1,\ldots ,4}\left(
    -\sta^*\nba\left(\Apat\PL+\Amat\PR\right)\,t\right.\nonumber\\
    &&\phantom{\sum_{a=1,2}\sum_{\alpha=1,\ldots ,4}}
    +\left.\sba^*\nba\left(\Apab\PL+\Amab\PR\right)\,b
  \right) \nonumber\\
   +&\mbox{\rm h.c.}&\ 
\end{eqnarray}
where $\Apmit,\ \Apmib,\ \Apmat,\ \Apmab$ are
\begin{equation}
\begin{array}{lll}
  \label{V1Apm}
  \Apit &=& \Rotc\Uo^*-\lt\Rttc\Ut^*\, ,\\
  \Amit &=& -\lb\Rotc\Vt\, ,\\
  \Apat &=& \Rotc\left(\Nt^*+\YL\tth\No^*\right)
                       +\sqrt{2}\lt\Rttc\Nth^*\, ,\\
  \Amat &=& \sqrt{2}\lt\Rotc\Nth
                       -\YRt\tth\Rttc\No\, ,\\
  \Apib &=& \Robc\Vo^*-\lb\Rtbc\Vt^*\, ,\\
  \Amib &=& -\lt\Robc\Ut\, ,\\
  \Apab &=& \Robc\left(\Nt^*-\YL\tth\No^*\right)
                       -\sqrt{2}\lb\Rtbc\Nf^*\, ,\\
  \Amab &=& -\sqrt{2}\lb\Robc\Nf
                       +\YRb\tth\Rtbc\No\, .
\end{array}
\end{equation}
with $\YL$ and $Y_R^{t,b}$ the weak hypercharges of the left-handed 
$SU(2)_L$ doublet and right-handed singlet fermion, and $\lt$ and 
$\lb$ the potentially significant Yukawa couplings -- Cf.  
eq.(\ref{eq:Yukawas}) -- normalized to the $SU(2)_L$ gauge coupling 
constant $g$.

\item{\bf quark--squark--gluino}
\begin{equation}
{\cal L}_{\sg q\tilde{q}}=- \frac{g_s}{\sqrt{2}}
\tilde q_{a,k}^{ *}\, \bar\sg_r 
\left(\lambda^r\right)_{kl} \left(R_{1a}^{(q)*} \PL - R_{2a}^{(q)*} 
\PR \right) q_l
+\mbox{ h.c.}
\label{eq:Lqsqglui}\end{equation}
where $\lambda^r$ are the Gell-Mann matrices.

\item{\bf squark--squark--Higgs}
\begin{equation}
  \label{LHsqsq}
  {\cal L}_{H^\pm\tilde{q}\tilde{q}}=\frac{g}{\sqrt{2}\mw}
%  H^+\sta^*R_{ca}^{*(t)}R_{db}^{(b)}g_{cd}\,\sbb\ +\ 
  H^-\sba^*G_{ab}\,\stb\ +\ 
  \mbox{\rm h.c.} 
\end{equation}
where we have introduced the coupling matrix
\begin{eqnarray}
  G_{ab}&=&R_{cb}^{(t)}R_{da}^{*(b)}\gccdd\label{Gdef}\nonumber\\
  \gccdd&=&\!\!\!\!\!\!\pmatrix{\!
    \mbs\tb+\mts\ctb-\mws\sin 2\beta & \!\!\!\!\!\!
    \mb\left(\mu+A_b\tb\right) \cr
    \mt\left(\mu+A_t\ctb\right) & 
    \!\!\!\!\!\!\mt\mb\left(\tb+\ctb\right)\!
    }\,.
    \label{gdef}
\end{eqnarray}
\item{\bf chargino--neutralino--charged Higgs}
\begin{equation}
  \label{LHcn}
  {\cal L}_{H^\pm\Psi^\mp\Psi^0}=
  -g\,H^-\nba\left(\HiaRc\PL+\HiaLc\PR\right)\cip\ +\
  \mbox{\rm h.c.} 
\end{equation}
with
\begin{equation}
\left\{
\begin{array}{lcl}
\QaiL&=&U_{i1}^*N_{\alpha 3}^*+
\frac{1}{\sqrt{2}}\left(N_{\alpha 2}^*+\tth N_{\alpha 1}^*
\right)U_{i2}^*\\
\QaiR&=&V_{i1}N_{\alpha 4}-
\frac{1}{\sqrt{2}}\left(N_{\alpha 2}+\tth N_{\alpha 1}
\right)V_{i2}\,.
\end{array}
\right.
\label{eq:QLQR}
\end{equation}
\end{itemize}

% \bibliography{tesi}

\end{document}